\documentclass[twocolumn,10pt]{article}
\usepackage{CSIMTA}
\usepackage[dvips]{graphicx}
\usepackage{amsmath}
\usepackage{amssymb}
\usepackage{natbib}

\begin{document}

\date{}

\title{A Mean-Field Model for Extended Stochastic Systems with Distributed
Time Delays}

\author{
Daniel Huber and Lev Tsimring \\ Institute for Nonlinear Science\\
University of California, San Diego\\ La Jolla, CA 92093-0402 \\ }

\maketitle
\thispagestyle{empty}

\noindent
{\bf\normalsize ABSTRACT}\newline {A network of noisy bistable
elements with global time-delayed couplings is considered. A
dichotomous mean field model has recently been developed describing
the collective dynamics in such systems with uniform time delays near
the bifurcation points. Here the theory is extended and applied to
systems with nonuniform time delays. For strong enough couplings the
systems exhibit delay-independent stationary states and
delay-dependent oscillatory states. We find that the regions of
oscillatory states in the parameter space are reduced with increasing
width of the time delay distribution function; that is, nonuniformity
of the time delays increases the stability of the trivial
equilibrium. However, for symmetric distribution functions the
properties of the oscillatory states depend only on the mean time
delay.}

\vspace{2ex}

\noindent
{\bf\normalsize KEY WORDS}\newline {Stochastic dynamics, mean-field
dynamics, delay differential equation, self-organization}

\section{Introduction}

The understanding of the collective dynamics in extended stochastic
systems with long range interactions is relevant for many domains in
physics, chemistry, biology and even social sciences. A popular and
effective generic model for such systems is the globally coupled
bistable-element-network, whose dynamical properties have been studied
in the absence \citep{Zanette97} and presence \citep[see][and
references therein]{Gammaitoni98} of noise and whose relevance for
critical phenomena
\citep{Dawson83}, spin systems \citep{Jung92}, neural networks
\citep{Jung92,Camperi98,Koulakov02}, 
genetic regulatory networks \citep{Gardner00} and decision making
processes in social systems \citep{Zanette97} has been pointed out.

In recent years it has been realized that time delays arising, for
example, from the finite propagation speed of signals are ubiquitous
in most physical and biological systems. The effects of the
delays on the behavior of various dynamical systems have been studied
\citep{Zanette00,Jeong02}, and some significant changes of the dynamical
properties have been demonstrated
\citep{Nakamura94,Bresseloff98,Choi85}. The effects of uniform time
delays in globally coupled networks of phase oscillators has been
explored by \citet{Yeung99}.
\citet{Tsimring01} studied the dynamics of a single 
noise activated, bistable element with time-delayed feedback.
Combining the properties of these two systems \citet{Huber03} studied
the cooperative dynamics of an ensemble of noisy bistable elements
with delayed couplings and a mean field theory based on
delay-differential master equations was developed.  However, the
theory assumed identical (i.e. uniform) time delays among all
elements. Although for many systems this approximation is
justified \citep{Salami03,Paulsson01}, most systems have distributed
coupling delays. Thus, we study in this paper the generalized case of
distributed time delays in a globally coupled network of bistable
elements.

\section{The model}

Our generalized model for the study of noise-activated, collective
dynamical phenomena in extended systems consists of $N$ Langevin
equations, each describing the overdamped noise driven motion of a
particle in a bistable potential $V=-x^2/2+x^4/4$, whose symmetry is
distorted by the time-delayed couplings to the other network elements,
\begin{equation}
\label{manyij}
\dot{x}_i=x_i-x^3_i+\frac{\varepsilon}{N}\sum\limits_{j=1}^N 
x_j(t-\tau_{ij})+\sqrt{2D}\xi(t),
\end{equation} 
where $\tau_{ij}$ are the time delays depending on the two coupled
elements $i$ and $j$. The strength of the feedback is $\varepsilon$
and $D$ denotes the variance of the Gaussian fluctuations $\xi(t)$,
which are $\delta$-correlated and mutually independent $\langle
\xi_i(t)\xi_j(t') \rangle = \delta(t-t')\delta_{ij}$.

We use an Euler method to explore model (\ref{manyij}) numerically
and focus our interest on the collective dynamics of the
bistable elements, i.e., on the dynamics of the mean field,
$X=N^{-1}\sum_{i=1}^N$. 

For $\varepsilon=0$, the elements are decoupled from each other. They
randomly and independently jump from one potential well to the other.
Therefore, in this case the mean field $X=0$. For small
$|\varepsilon|$, the mean field remains zero. However, for a strong
enough feedback the system undergoes ordering transitions and
demonstrates multistability.  That is, for a strong enough positive
coupling the systems undergoes a pitchfork bifurcation and adopts a
non-zero stationary mean field $X>0$, and transitions to a variety of
stable oscillatory mean field states via Hopf bifurcations, are
observed for strong enough positive and negative feedbacks.

In the general case, model (\ref{manyij}), in which the time delays
depend on both the ``transmitting'' and the ``receiving'' element, cannot
directly be described in terms of a mean field theory. However, the
system becomes mathematically tractable if we assume that the time delay
does only depend on the ``transmitting'' elements $j$,
\begin{equation}
\label{manyj}
\dot{x}_i=x_i-x^3_i+\frac{\varepsilon}{N}\sum\limits_{j=1}^N 
x_j(t-\tau_{j})+\sqrt{2D}\xi(t).
\end{equation}

\begin{figure}
\centerline{\includegraphics[width=7cm]{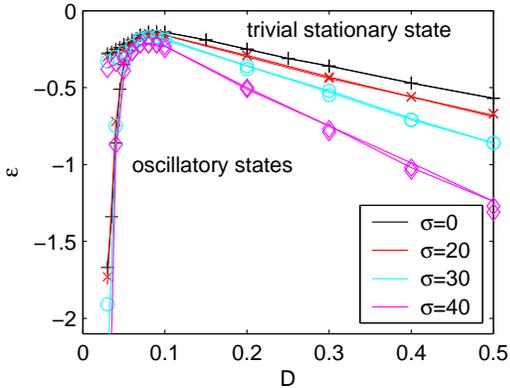}}
\caption{\label{mixgaussphase} The critical coupling of the Hopf bifurcation
as a function of the noise strength $D$ for different $\sigma$
of the Gaussian time delay distribution with $\bar{\tau}=100$. The
markers and the solid lines depict the critical couplings resulting
from model (\ref{manyij}) and model (\ref{manyj}), respectively.}
\end{figure}

In order to check if such a simplification is justified, numerical
simulations of model (\ref{manyij}) and (\ref{manyj}) are carried out
and compared. In these simulations the distribution of the time delays
is Gaussian, i.e., it is fully determined by its mean $\bar{\tau}$ and
variance $\sigma$. Fig. \ref{mixgaussphase}, which compares the
critical coupling strength of the Hopf bifurcation for different
$\sigma$, suggests that the above simplification is justified
in order to study the stability properties of a
bistable-element-network with time delays.

This surprising result not only renders possible an analytical
description of networks with distributed delays but also implicates
that the number of operations, which have to be carried out to study
such systems numerically, can be reduced from ${\cal O}(N^2)$ to
${\cal O}(N)$.

\section{Dichotomous Theory}

In this section we derive the dichotomous theory for the description
of globally coupled bistable-element-networks with distributed time
delays.

The dichotomous theory, valid in the limit of small coupling strength
and small noise, neglects intra-well fluctuations of $x_i$. Thus, each
bistable element can be replaced by a discrete two-state system which
can only take the values $s_{1,2}=\pm1$. Then the collective dynamics
of the entire system is described by the master equations for the
occupation probabilities of these states $n_{1,2}$. This approach has
been successfully used in studies of stochastic and coherence
resonance
\citep[e.g.][]{McNamara89,Gammaitoni98,Jung92,Tsimring01}, 
and \citet{Huber03} used it to study the special case of a network
with uniform time delays. They found that while away from the
transition points the system dynamics is well described by a Gaussian
approximation, near the bifurcation points a description in terms of a
dichotomous theory is more adequate.

In order to apply the dichotomous theory to a network with distributed
time delays, we coarse grain system (\ref{manyj}). The coarse graining
is accomplished as follows: The range of possible time delays is
divided up in $M$ intervals $I_k\;\{k=1,2,\ldots,M\}$. The size of the
intervals $\Delta_k$ is chosen, so that the number of bistable
oscillators associated with a delay fitting in a particular interval,
is for each interval the same $m=N/M$. In this way oscillator groups
are formed whose mean field can be expressed as,
\begin{equation}
\Omega_k(t)\equiv\frac{1}{m}
\sum_{\tau_j\in I_k} x_j(t),
\end{equation}
where $I_k\equiv[\tau_k,\tau_{k+1}[$, $\tau_k=\sum_{l=1}^{k-1}\Delta_l$
and $j=1\ldots N$.

Assuming that $\Delta_k\ll\bar{\tau}/\sigma$, where $\bar{\tau}$ and
$\sigma$ are the mean and the variance of the time delay distribution,
Eq. (\ref{manyj}) can then be approximated by,
\begin{equation}
\label{omegadyn}
x_i=x_i-x_i^3+\frac{\varepsilon}{M}\sum\limits_{k=1}^M \Omega_k(t-\tau_k)
+\sqrt{2D}\xi(t).
\end{equation}

The dynamics of a single element $x_i$ is determined by the hopping
rates $p_{12}$ and $p_{21}$, i.e., by the probabilities to hop over
the potential barrier from $s_1$ to $s_2$ and from $s_2$ to $s_1$,
respectively. In a globally coupled system, in which the time delays
depend only on the transmitting elements, $p_{12}$ and $p_{21}$ are
identical for all elements and the master equations expressing the
dynamics of Eq. (\ref{omegadyn}) in terms of occupation probabilities
read,
\begin{eqnarray}
\label{ndot1}
\dot{n}_{1,k}  &=&  -p_{12}n_{1,k}+p_{21}n_{2,k}\\
\dot{n}_{2,k}  &=&  p_{12}n_{1,k}-p_{21}n_{2,k}.
\end{eqnarray} 
The hopping probabilities $p_{12,21}$ are given by Kramers' transition
rate \citep{Kramers40} for the instantaneous potential well, which for
our system in the limit of small noise $D$ and coupling strength
$\varepsilon$ reads \citep[cf.][]{Tsimring01},
\begin{equation}
p_{12,21}=\frac{\sqrt{2\mp 3\alpha}}{2\pi}
\exp\left(-\frac{1\mp4\alpha}{4 D}\right),
\end{equation} 
where $\alpha=(\varepsilon/M)\sum_{k=1}^M \Omega_k (t-\tau_k)$.

For large oscillator groups ($m\to\infty$),
$\Omega_k=n_{1,k}s_1+n_{2,k}s_2=n_{2,k}-n_{1,k}$ and
$n_{1,k}+n_{2,k}=1$ holds. With these terms, we can find the following
set of equations:
\begin{equation}
\label{manyhop}
\dot \Omega_k(t) = p_{12}-p_{21}-(p_{21}+p_{12})\Omega_k(t).
\end{equation}
The Jacobian matrix of this system is given through,
\begin{equation}
J=c
\begin{pmatrix}
a_1+b & a_2 & \dots & a_M \\
a_1 & a_2+b & \dots & a_M \\
\hdotsfor{4}              \\
a_1 & a_2 & \dots & a_M+b \\
\end{pmatrix},
\end{equation}  
where $c=-\sqrt{2}\exp(-1/4D)/(4M\pi D)$, 
$a_k=\varepsilon(3D-4)\exp(-\lambda\tau_k)$ and $b=4MD$.
With this Jacobian the characteristic equation, determining
the stability of the trivial equilibrium $X=0$, becomes
\begin{equation}
\label{ceq1}
(bc-\lambda)^{M-1}\left(c\left[b+\sum_{j=1}^M a_j\right]-\lambda\right)=0.
\end{equation}
The trivial equilibrium loses its stability and undergoes a pitchfork
bifurcation, describing the transition to a steady nonzero mean field,
when the real solutions of the characteristic equation (i.e. the
eigenvalues of the Jacobian) become positive. Thus setting $\lambda=0$
and solving Eq. (\ref{ceq1}) for $\varepsilon$ yields the critical
coupling for the pitchfork instability,
\begin{equation}
\label{epspf}
\varepsilon_{\rm p}=\frac{4D}{4-3D}.
\end{equation}
A Hopf bifurcation indicating the transition to an oscillatory mean
field state occurs when the real part of the complex eigenvalues
becomes positive. Therefore, the properties of the corresponding
instabilities (i.e. frequencies and coupling strengths at the
bifurcation points) can be found by substituting $\lambda=\mu+{\rm
i}\omega$ into Eq. (\ref{ceq1}), separating real and imaginary parts
and setting $\mu=0$. For the frequencies of the unstable modes we
find,
\begin{equation} 
\label{manyomega}
\omega\bar{\tau} = -\frac{\sqrt{2}}{\pi}
\exp(-1/4D)\bar{\tau}\frac{I_{\rm s}}{I_{\rm c}},
\end{equation}
where 
\begin{equation}
I_{\rm s}=\frac{1}{M}\sum_{k=1}^{M}\sin\omega\tau_k,\;\:
I_{\rm c}=\frac{1}{M}\sum_{k=1}^{M}\cos\omega\tau_k.  
\end{equation}
For large systems $N\to\infty$, the number of groups $M\to\infty$
and thus
\begin{equation}
\label{integrals}
I_s=\int\limits_0^{\infty}P(\tau)\sin\omega\tau d\tau,\;\:
I_c=\int\limits_0^{\infty}P(\tau)\cos\omega\tau d\tau,
\end{equation} 
where $P(\tau)$ is the time delay distribution function.

We can express the time delay distribution function in terms of
cumulant moments $\kappa_n$ \citep{Kampen03} and solve the integrals in
(\ref{integrals}):
\begin{equation}
I_s=\sin(g_1)\exp({g_2}),\;\:
I_c=\cos(g_1)\exp({g_2}),
\end{equation}
where
\begin{eqnarray}
g_1 & = & \sum_{m=0}^\infty \frac{({\rm i}\omega)^{2m+1}}
{{\rm i}(2m+1)!}\kappa_{2m+1},\\ 
g_2 & = & \sum_{m=1}^\infty\frac{({\rm i}\omega)^{2m}}{(2m)!}\kappa_{2m}.
\end{eqnarray}
Consequently, 
\begin{equation}
\label{tang1}
\frac{I_{\rm s}}{I_{\rm c}}=\tan(g_1).
\end{equation}
Since for symmetric distribution functions all odd cumulant moments
except the first one $\kappa_1=\bar{\tau}$ are zero, $I_{\rm s}/I_{\rm
c}=\tan\omega\bar{\tau}$ holds. That is, in the case of a symmetric
distribution of the time delays, the frequencies of the unstable modes
in Eq. (\ref{manyomega}) depend only on the mean time delay.

Let us now determine the critical coupling of the Hopf bifurcation.
For large time delays $\bar{\tau}\gg\tau_k$ ($\tau_k$ is the inverse
Kramers escape rate from one well into the other) the low-order
solutions of the transcendent equation (\ref{manyomega}) yield
frequencies $\omega\ll 1$. Thus the real part of equation (\ref{ceq1})
can be linearized near $\omega=0$ and the critical coupling of the
Hopf bifurcation becomes,
\begin{equation} 
\varepsilon_{\rm H}= \frac{4 D\pi\omega}
{(3D-4)\left(\frac{1}{N}\sqrt{2}\exp(-1/4D)I_{\rm s}
-\left[1-\frac{1}{N}\right]\pi\omega I_{\rm c}\right)}.
\end{equation} 
Then, for large systems $N\to\infty$ the critical coupling is,
\begin{equation}
\label{epsosc}
\varepsilon_{\rm H}= \frac{4 D}
{(4-3D)I_{\rm c}},
\end{equation}
with $I_{\rm c}=3\sin(\omega\bar{\tau})\sin(5\omega\sigma/3)/
(5\omega\sigma)$ and $I_{\rm c}
=\cos(\omega\bar{\tau})\exp(-\omega^2\sigma^2/2)$ for a
uniform and a Gaussian distribution, respectively.

Eq. (\ref{manyomega}) and (\ref{epsosc}) have a multiplicity of solutions
indicating that multistability occurs in the system beyond a certain
coupling strength.

\section{Results}

\begin{figure}
\centerline{\includegraphics[width=8.3cm]{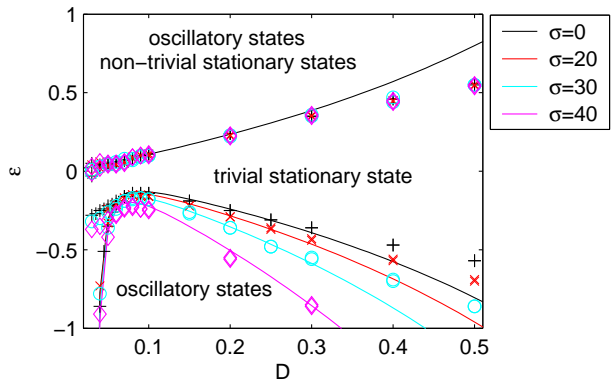}}
\caption{\label{complantheo} Phase diagram of the globally coupled
bistable-element-network with uniformly distributed time delays
derived from the theoretical model (solid lines) and numerical
simulations of the Langevin model (markers). The phase diagram is
shown for different standard deviations $\sigma$ of the delay
distribution function.  The mean time delay is $\bar{\tau}=100$.}
\end{figure}

\begin{figure}
\centerline{\includegraphics[width=7cm]{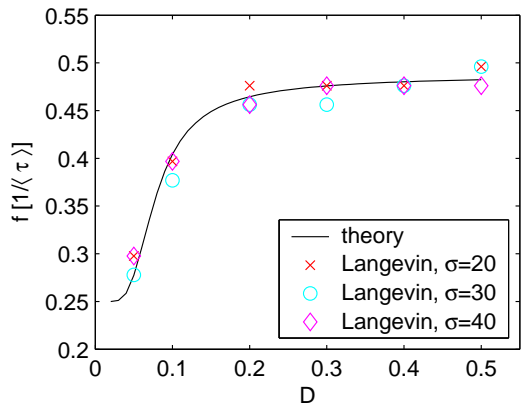}}
\caption{\label{fgeneral} The frequencies of the unstable modes at
the bifurcation points resulting from the Langevin model (markers) and
the the dichotomous model (solid line). For symmetric distributions
the frequencies depend only on the mean time delay (see
Eq. \ref{manyomega} and \ref{tang1}).}
\end{figure}

Eq. (\ref{epspf}), (\ref{manyomega}) and (\ref{epsosc}) are used to
determine the phase diagram and the frequencies of the unstable
oscillatory modes $f=\omega/(2\pi)$ of a bistable-element-network with
uniformly distributed time delays\footnote{This should not be confused
with uniform time delays, which means that the delay for each coupling
is the same.}. The theoretical predictions are verified with numerical
simulations of the Langevin model (\ref{manyij}). The results are
shown in Fig. \ref{complantheo} and Fig. \ref{fgeneral}.

The figures show that near the bifurcation points the predictions by
the dichotomous theory are reasonably good for small noise
$(D\lesssim 0.3)$ and consequently in this regime the Langevin
models (\ref{manyij}) and (\ref{manyj}) are dynamically
equivalent. 

Fig. \ref{complantheo} also shows that the regions in the phase space
where mean field oscillations occur are reduced with increasing width
$\sigma$ of the time delay distribution function, meaning that
nonuniformity of the time delays inhibits the occurence of Hopf
bifurctions, i.e., increases the stability of the trivial equilibrium.



\begin{figure*}
\centerline{\includegraphics[width=13cm]{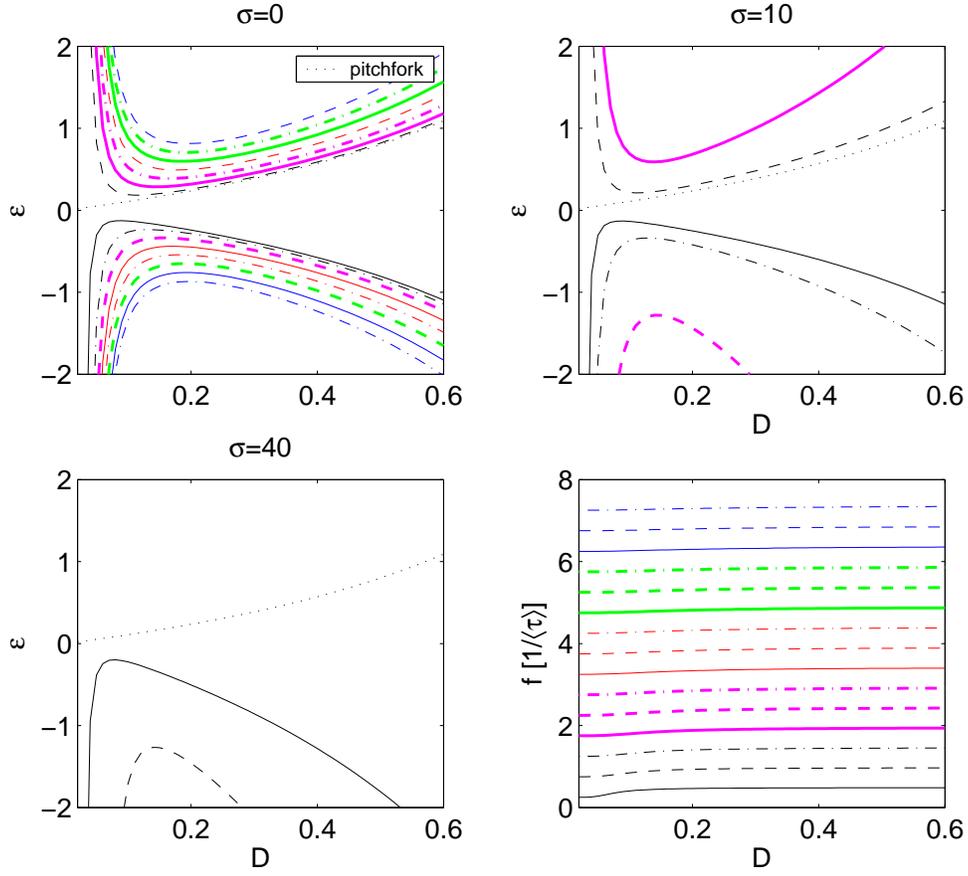}}
\caption{\label{multi} Upper panels and lower left panel: 
Phase diagrams of systems with uniformly distributed time delays,
where $\bar{\tau}=100$ and $\sigma=0,\;10,\;40$. The dotted line
depicts the critical coupling of the pitchfork bifurcation and the
other lines depict those of the primary Hopf bifurcation as well as
some higher order solutions (i.e. solution 1-16) of
Eq. (\ref{manyomega}) and (\ref{epsosc}). Lower right panel: The
frequencies of the corresponding unstable modes which do not depend on
$\sigma$, but slightly vary with the noise strength $D$.}
\end{figure*}

In Fig. \ref{multi} the bifurcation diagrams including higher order
solutions of Eq. (\ref{manyomega}) and (\ref{epsosc}) are
presented. This Figure shows that for a strong enough positive
$(\varepsilon=\varepsilon_{\rm p})$ coupling the trivial equilibrium
loses its stability via a pitchfork bifurcation while for a strong
enough negative coupling $(\varepsilon=\varepsilon^1_{\rm H-})$ a Hopf
bifurcation determined by the primary solution of
Eq. (\ref{manyomega}) and (\ref{epsosc}) occurs. The higher order
solutions of these equations provide the multistability of the
system. For a positive feedback several oscillatory states with
frequencies $f_k\approx k/\bar{\tau}$ are observed for
$\varepsilon>\varepsilon^{k}_{\rm H+}\;\;\{k=1,2,\ldots\}$. The
transition points are ordered as follows, $0<\varepsilon_{\rm
st}<\varepsilon^{1}_{\rm H+}<\varepsilon^{2}_{\rm H+}\ldots$ If the
feedback is negative, the system has oscillatory solutions with
periods $f_l\approx (2l+1)/(2\bar{\tau})$ for
$\varepsilon<\varepsilon^{l}_{\rm H-}\;\;\{l=0,1,\ldots\}$, where
$0>\varepsilon^{0}_{\rm H-}>\varepsilon^{1}_{\rm H-}\ldots$ In the
above annotation a (+/-) index means that the corresponding value is
associated with a negative and positive feedback, respectively.

\section{Summary and conclusions}

We generalized a dichotomous theory based on delay-differential
master-equations to account for the dynamics of globally coupled
networks of bistable elements with nonuniform time-delays. As in the
case of uniform time delays these systems possess a nonzero stationary
mean field for a strong enough positive feedback whose properties are
time delay independent and a multiplicity of time delay dependent
stable oscillatory states for both positive and negative feedback.

For symmetric time delay distributions the frequencies of the
oscillations depend only on the mean time delay.  However, the
critical couplings of the corresponding Hopf bifurcations depend also
on the width of the time delay distribution; that is, the critical
coupling strengths and consequently the stability of the trivial
equilibrium are increased for broadly distributed time delays. This
may be important for time delay systems such as neural networks and
genetic regulatory networks, since the degree of time delay
nonuniformity, which is often related to the diversity in the
connectivity of the underlying network, affects the accessibility of
the nontrivial dynamical states.

The bifurcations of the trivial equilibrium are well described by the
dichotomous theory in the limit of small noise and coupling strength.
Far away from the transition points a theoretical description of the
mean field dynamics can be found using a Gaussian approximation
\citep{Desai78,Huber03}. However, a theoretical approach for the
description of the dynamics in the regime of strong noise near the
bifurcation points is still lacking.

This paper discusses the dynamics of globally coupled systems with
time delays. It is assumed that all elements are coupled with uniform
(i.e. identical) strength $\varepsilon$.  However, many networks have
sparse couplings and nonuniform coupling strength, which may endow the
system with a complexer dynamics. This issue will be discussed in an
upcoming paper.

\subsubsection*{Acknowledgments} This work was supported by the Swiss National
Science Foundation (D.H.) and by the U.S. Department of Energy, Office
of Basic Energy Sciences under grant DE-FG-03-96ER14592 (L.T.).


\begin{thebibliography}{21}
\expandafter\ifx\csname natexlab\endcsname\relax\def\natexlab#1{#1}\fi
\expandafter\ifx\csname bibnamefont\endcsname\relax
  \def\bibnamefont#1{#1}\fi
\expandafter\ifx\csname bibfnamefont\endcsname\relax
  \def\bibfnamefont#1{#1}\fi
\expandafter\ifx\csname citenamefont\endcsname\relax
  \def\citenamefont#1{#1}\fi
\expandafter\ifx\csname url\endcsname\relax
  \def\url#1{\texttt{#1}}\fi
\expandafter\ifx\csname urlprefix\endcsname\relax\def\urlprefix{URL }\fi
\providecommand{\bibinfo}[2]{#2}
\providecommand{\eprint}[2][]{\url{#2}}

\bibitem[{\citenamefont{{Zanette}}(1997)}]{Zanette97}
\bibinfo{author}{\bibfnamefont{D.~H.} \bibnamefont{{Zanette}}},
  \bibinfo{journal}{Phys. Rev. E} \textbf{\bibinfo{volume}{55}},
  \bibinfo{pages}{5315} (\bibinfo{year}{1997}).

\bibitem[{\citenamefont{{Gammaitoni} et~al.}(1998)\citenamefont{{Gammaitoni},
  {H{\" a}nggi}, {Jung}, and {Marchesoni}}}]{Gammaitoni98}
\bibinfo{author}{\bibfnamefont{L.}~\bibnamefont{{Gammaitoni}}},
  \bibinfo{author}{\bibfnamefont{P.}~\bibnamefont{{H{\" a}nggi}}},
  \bibinfo{author}{\bibfnamefont{P.}~\bibnamefont{{Jung}}}, \bibnamefont{and}
  \bibinfo{author}{\bibfnamefont{F.}~\bibnamefont{{Marchesoni}}},
  \bibinfo{journal}{Rev. Mod. Phys.} \textbf{\bibinfo{volume}{70}},
  \bibinfo{pages}{223} (\bibinfo{year}{1998}).

\bibitem[{\citenamefont{{Dawson}}(1983)}]{Dawson83}
\bibinfo{author}{\bibfnamefont{D.}~\bibnamefont{{Dawson}}},
  \bibinfo{journal}{J. Stat. Phys.} \textbf{\bibinfo{volume}{29}},
  \bibinfo{pages}{31} (\bibinfo{year}{1983}).

\bibitem[{\citenamefont{{Jung} et~al.}(1992)\citenamefont{{Jung}, {Behn},
  {Pantazelou}, and {Moss}}}]{Jung92}
\bibinfo{author}{\bibfnamefont{P.}~\bibnamefont{{Jung}}},
  \bibinfo{author}{\bibfnamefont{U.}~\bibnamefont{{Behn}}},
  \bibinfo{author}{\bibfnamefont{E.}~\bibnamefont{{Pantazelou}}},
  \bibnamefont{and} \bibinfo{author}{\bibfnamefont{F.}~\bibnamefont{{Moss}}},
  \bibinfo{journal}{Phys. Rev. A} \textbf{\bibinfo{volume}{46}},
  \bibinfo{pages}{R1709} (\bibinfo{year}{1992}).

\bibitem[{\citenamefont{{Camperi} and {Wang}}(1998)}]{Camperi98}
\bibinfo{author}{\bibfnamefont{M.}~\bibnamefont{{Camperi}}} \bibnamefont{and}
  \bibinfo{author}{\bibfnamefont{X.}~\bibnamefont{{Wang}}},
  \bibinfo{journal}{J. Comp. Neurosci.} \textbf{\bibinfo{volume}{5}},
  \bibinfo{pages}{383} (\bibinfo{year}{1998}).

\bibitem[{\citenamefont{{Koulakov} et~al.}(2002)\citenamefont{{Koulakov},
  {Raghavachari}, {Kepecs}, and {Lisman}}}]{Koulakov02}
\bibinfo{author}{\bibfnamefont{A.}~\bibnamefont{{Koulakov}}},
  \bibinfo{author}{\bibfnamefont{S.}~\bibnamefont{{Raghavachari}}},
  \bibinfo{author}{\bibfnamefont{A.}~\bibnamefont{{Kepecs}}}, \bibnamefont{and}
  \bibinfo{author}{\bibfnamefont{J.}~\bibnamefont{{Lisman}}},
  \bibinfo{journal}{Nat. Neurosci.} \textbf{\bibinfo{volume}{5}},
  \bibinfo{pages}{775} (\bibinfo{year}{2002}).

\bibitem[{\citenamefont{{Gardner} et~al.}(2000)\citenamefont{{Gardner},
  {Cantor}, and {Collins}}}]{Gardner00}
\bibinfo{author}{\bibfnamefont{T.}~\bibnamefont{{Gardner}}},
  \bibinfo{author}{\bibfnamefont{C.}~\bibnamefont{{Cantor}}}, \bibnamefont{and}
  \bibinfo{author}{\bibfnamefont{J.}~\bibnamefont{{Collins}}},
  \bibinfo{journal}{Nature} \textbf{\bibinfo{volume}{403}},
  \bibinfo{pages}{339} (\bibinfo{year}{2000}).

\bibitem[{\citenamefont{{Zanette}}(2000)}]{Zanette00}
\bibinfo{author}{\bibfnamefont{D.~H.} \bibnamefont{{Zanette}}},
  \bibinfo{journal}{Phys. Rev. E} \textbf{\bibinfo{volume}{62}},
  \bibinfo{pages}{3167} (\bibinfo{year}{2000}).

\bibitem[{\citenamefont{{Jeong} et~al.}(2002)\citenamefont{{Jeong}, {Ko}, and
  {Moon}}}]{Jeong02}
\bibinfo{author}{\bibfnamefont{S.}~\bibnamefont{{Jeong}}},
  \bibinfo{author}{\bibfnamefont{T.}~\bibnamefont{{Ko}}}, \bibnamefont{and}
  \bibinfo{author}{\bibfnamefont{H.}~\bibnamefont{{Moon}}},
  \bibinfo{journal}{Phys. Rev. Lett.} \textbf{\bibinfo{volume}{89}},
  \bibinfo{pages}{154104} (\bibinfo{year}{2002}).

\bibitem[{\citenamefont{{Nakamura} et~al.}(1994)\citenamefont{{Nakamura},
  {Tominaga}, and {Munakata}}}]{Nakamura94}
\bibinfo{author}{\bibfnamefont{Y.}~\bibnamefont{{Nakamura}}},
  \bibinfo{author}{\bibfnamefont{F.}~\bibnamefont{{Tominaga}}},
  \bibnamefont{and}
  \bibinfo{author}{\bibfnamefont{T.}~\bibnamefont{{Munakata}}},
  \bibinfo{journal}{Phys. Rev. E} \textbf{\bibinfo{volume}{49}},
  \bibinfo{pages}{4849} (\bibinfo{year}{1994}).

\bibitem[{\citenamefont{{Bressloff} and {Coombes}}(1998)}]{Bresseloff98}
\bibinfo{author}{\bibfnamefont{P.~C.} \bibnamefont{{Bressloff}}}
  \bibnamefont{and}
  \bibinfo{author}{\bibfnamefont{S.}~\bibnamefont{{Coombes}}},
  \bibinfo{journal}{Phys. Rev. Lett.} \textbf{\bibinfo{volume}{80}},
  \bibinfo{pages}{4815} (\bibinfo{year}{1998}).

\bibitem[{\citenamefont{{Choi} and {Huberman}}(1985)}]{Choi85}
\bibinfo{author}{\bibfnamefont{M.~Y.} \bibnamefont{{Choi}}} \bibnamefont{and}
  \bibinfo{author}{\bibfnamefont{B.~A.} \bibnamefont{{Huberman}}},
  \bibinfo{journal}{Phys. Rev. B} \textbf{\bibinfo{volume}{31}},
  \bibinfo{pages}{2862} (\bibinfo{year}{1985}).

\bibitem[{\citenamefont{{Yeung} and {Strogatz}}(1999)}]{Yeung99}
\bibinfo{author}{\bibfnamefont{M.~K.~S.} \bibnamefont{{Yeung}}}
  \bibnamefont{and} \bibinfo{author}{\bibfnamefont{S.~H.}
  \bibnamefont{{Strogatz}}}, \bibinfo{journal}{Phys. Rev. Lett.}
  \textbf{\bibinfo{volume}{82}}, \bibinfo{pages}{648} (\bibinfo{year}{1999}).

\bibitem[{\citenamefont{{Tsimring} and {Pikovsky}}(2001)}]{Tsimring01}
\bibinfo{author}{\bibfnamefont{L.~S.} \bibnamefont{{Tsimring}}}
  \bibnamefont{and}
  \bibinfo{author}{\bibfnamefont{A.}~\bibnamefont{{Pikovsky}}},
  \bibinfo{journal}{Phys. Rev. Lett.} \textbf{\bibinfo{volume}{87}},
  \bibinfo{pages}{250602} (\bibinfo{year}{2001}).

\bibitem[{\citenamefont{{Huber} and {Tsimring}}(2003)}]{Huber03}
\bibinfo{author}{\bibfnamefont{D.}~\bibnamefont{{Huber}}} \bibnamefont{and}
  \bibinfo{author}{\bibfnamefont{L.~S.} \bibnamefont{{Tsimring}}},
  \bibinfo{journal}{Phys. Rev. Lett.} \textbf{\bibinfo{volume}{91}},
  \bibinfo{pages}{260601} (\bibinfo{year}{2003}).

\bibitem[{\citenamefont{{Salami} et~al.}(2003)\citenamefont{{Salami}, {Itami},
  {Tsumoto}, and {Kimura}}}]{Salami03}
\bibinfo{author}{\bibfnamefont{M.}~\bibnamefont{{Salami}}},
  \bibinfo{author}{\bibfnamefont{C.}~\bibnamefont{{Itami}}},
  \bibinfo{author}{\bibfnamefont{T.}~\bibnamefont{{Tsumoto}}},
  \bibnamefont{and} \bibinfo{author}{\bibfnamefont{F.}~\bibnamefont{{Kimura}}},
  \bibinfo{journal}{PNAS} \textbf{\bibinfo{volume}{100}}, \bibinfo{pages}{6174}
  (\bibinfo{year}{2003}).

\bibitem[{\citenamefont{{Paulsson} and {Ehrenberg}}(2001)}]{Paulsson01}
\bibinfo{author}{\bibfnamefont{J.}~\bibnamefont{{Paulsson}}} \bibnamefont{and}
  \bibinfo{author}{\bibfnamefont{M.}~\bibnamefont{{Ehrenberg}}},
  \bibinfo{journal}{Q. Rev. Biophys.} \textbf{\bibinfo{volume}{34}},
  \bibinfo{pages}{1} (\bibinfo{year}{2001}).

\bibitem[{\citenamefont{{McNamara} and {Wiesenfeld}}(1989)}]{McNamara89}
\bibinfo{author}{\bibfnamefont{B.}~\bibnamefont{{McNamara}}} \bibnamefont{and}
  \bibinfo{author}{\bibfnamefont{K.}~\bibnamefont{{Wiesenfeld}}},
  \bibinfo{journal}{Phys. Rev. A} \textbf{\bibinfo{volume}{39}},
  \bibinfo{pages}{4854} (\bibinfo{year}{1989}).

\bibitem[{\citenamefont{{Kramers}}(1940)}]{Kramers40}
\bibinfo{author}{\bibfnamefont{H.}~\bibnamefont{{Kramers}}},
  \bibinfo{journal}{Physica (Utrecht)} \textbf{\bibinfo{volume}{7}},
  \bibinfo{pages}{284} (\bibinfo{year}{1940}).

\bibitem[{\citenamefont{{Van Kampen}}(2003)}]{Kampen03}
\bibinfo{author}{\bibfnamefont{N.}~\bibnamefont{{Van Kampen}}},
  \emph{\bibinfo{title}{Stochastic Processes in Physics and Chemistry}}
  (\bibinfo{publisher}{Elesiver Scinece B.V.}, \bibinfo{address}{Amsterdam, The
  Netherlands}, \bibinfo{year}{2003}).

\bibitem[{\citenamefont{{Desai} and {Zwanzig}}(1978)}]{Desai78}
\bibinfo{author}{\bibfnamefont{R.~C.} \bibnamefont{{Desai}}} \bibnamefont{and}
  \bibinfo{author}{\bibfnamefont{R.}~\bibnamefont{{Zwanzig}}},
  \bibinfo{journal}{J. Stat. Phys.} \textbf{\bibinfo{volume}{19}},
  \bibinfo{pages}{1} (\bibinfo{year}{1978}).

\end{thebibliography}

\end{document}